\documentclass[12pt]{article}
\usepackage{fleqn,amssymb,amsmath,amscd,epsfig,version,theorem} %

\newcommand{\Section}[1]{\section{#1}\setcounter{equation}{0}}
\newtheorem{theorem}{Theorem} [section]
\newtheorem {lemma}[theorem]{Lemma}
\newtheorem {proposition}[theorem]{Proposition}
\newtheorem {corollary}[theorem]{Corollary}
\theorembodyfont{\normalfont}

\newtheorem {conjecture}[theorem]{Conjecture}

\newtheorem {examples}[theorem]{Examples}

\newtheorem {remark}[theorem]{Remark}
\newtheorem {remarks}[theorem]{Remarks}
\newcommand{\beq}{\begin{equation}}
\newcommand{\eeq}{\end{equation}}
\newcommand{\Leq}[1]{\label{#1}\end{equation}}
\newcommand{\beqn}{\begin{eqnarray}}
\newcommand{\eeqn}{\end{eqnarray}}
\newcommand{\beqno}{\begin{eqnarray*}}
\newcommand{\eeqno}{\end{eqnarray*}}
\newcommand{\es}{\emptyset}
\renewcommand {\l}{\left}
\newcommand {\ri}{\right}
\newcommand {\vep}{\varepsilon}

\newcommand {\LA}{\left\langle}
\newcommand {\RA}{\right\rangle}

\newcommand {\eh}{{\textstyle \frac{1}{2}}}

\newcommand {\ar}{\rightarrow}
\newcommand {\sign}{{\rm sign}}

\newcommand {\tr}{{\rm tr}}

\newcommand {\bC}{{\mathbb C}}

\newcommand {\bN}{{\mathbb N}}
\newcommand {\bR}{{\mathbb R}}
\newcommand {\bZ}{{\mathbb Z}}
\newcommand {\bT}{{\mathbb T}}

\newcommand{\idty}{{\rm 1\mskip-4mu l}} 
\newcommand{\cD}{{\cal D}} %
\newcommand{\cE}{{\cal E}} %

\newcommand{\cO}{{\cal O}} 
\newcommand{\cP}{{\cal P}}
\newcommand{\cN}{{\cal N}}

\newcommand{\ov}{\overline}
\newcommand{\bem}{\l(\! \begin{array}}
\newcommand{\eem}{\end{array}\!\ri)}
\newcommand{\bsm}{\left(\begin{smallmatrix}} 
\newcommand{\esm}{\end{smallmatrix}\right)}  

\newcommand{\hp}{{\hat{p}}}   %

\newcommand{\hsi}{{\hat{\sigma}}}
\newcommand{\hq}{{\hat{q}}}   %

\newcommand{\qmbox}[1]{\quad\mbox{#1}\quad}
\renewcommand {\max}{{{\rm max}}}

\newcommand{\diag}{{\rm diag}}

\newcommand{\av}[1]{\left\langle#1\right\rangle}
\newcommand{\V}{V}
\newcommand{\St}{{S_t}}
\newcommand{\Ct}{{{\cal C}_t}}

\newcommand{\Dt}{{\cal D}_t}
\newcommand{\brac}[1]{\LA #1\RA}
\newcommand{\LL}{\cal{L}}
\begin{document}
\title{On the Form Factor for the Unitary Group}

\author{
Mirko Degli Esposti\thanks{
        Dipartimento di Matematica, Universit\`a di Bologna,
        Piazza di Porta S. Donato, 5, I-40127 Bologna, Italy,
        e-mail: desposti@dm.unibo.it}
\and
Andreas Knauf\thanks{Mathematisches Institut der
        Universit\"at Erlangen-N\"urnberg. Bismarckstr. 1 1/2,
        D-91054 Erlangen, Germany.
        e-mail: knauf@mi.uni-erlangen.de }}
\date{November 3, 2003}
\maketitle
\begin{abstract}\noindent
We study the combinatorics of the contributions to the form factor of the
group $U(N)$ in the large $N$ limit. This relates to questions about
semiclassical contributions to the form factor of quantum systems described by
the unitary ensemble.
\end{abstract}

\tableofcontents
\Section{Introduction}
%
The {\em form factor} associated to a self--adjoint operator $H$ is a real--valued
function describing statistical properties of its spectrum. For sake of
simplicity we assume that $H$ acts on finite--dimensional Hilbert space and
thus has eigenvalues $E_1,\ldots,E_N\in\bR$. Then we consider the Fourier
transform of the measure
\[\mu:=\frac{1}{N}\sum_{j,k=1}^N\delta_{E_j-E_k}\]
and obtain the form factor
\[K(t):=\int_\bR\exp(-itE)\,d\mu(E) = \frac{1}{N}\big|\tr(U(t))\big|^2,\]
with the unitary time evolution $U(t):=\exp(-iHt)$ generated by $H$.

It is an empirical fact and a physical conjecture (see Bohigas,
Giannoni and Schmit \cite{BGS} and also \cite{Ha}) that most form
factors encountered in physical quantum systems resemble the form
factor associated to a so--called random matrix ensemble (see
Mehta \cite{Me}).

The simplest of these is the so--called {\em unitary ensemble} on which we shall
concentrate below.
This is given by the unitary group $U(N)$ equipped with Haar probability
measure $\mu_N$. Its form factor is defined as
\beq
K_N(t):=\frac{1}{N}\LA|\tr(U^t)|^2\RA_N \qquad (t\in\bZ),
\Leq{KN1}
with the expectation $\LA f\RA\equiv\LA f\RA_N:=\int_{U(N)}f\, d\mu_N$ of a
continuous function $f:U(N)\to\bC$.

As the map $U\mapsto\tr(U^t)$ is a class function on the unitary group, we
can apply Weyl's integration formula
\beq
\int_{U(N)}f\,d\mu_N=\frac{1}{N!}\int_{\bT^N}f\Delta^2\,d\nu_N
\Leq{UN}
to evaluate (\ref{KN1}). In (\ref{UN}) $f:U(N)\to\bC$ is assumed to be a class
function. $\bT^N\subseteq U(N)$ is a maximal torus and may be identified
with the subgroup of diagonal matrices. $d\nu_N$ denotes Haar measure on
$\bT^N$. Finally for $h:=\diag(h_1,\ldots,h_N)\in\bT^n\subseteq U(N)$
\[\Delta(h):=\sum_{1\leq j<k\leq N}|h_j-h_k|\]
is the modulus of Vandermonde's determinant for $h_1,\ldots,h_N$. The
combinatorial factor $N!$ is the order of the symmetric group $S_N$ making
its appearance as the Weyl group, see e.g. Fulton and Harris \cite{FH}.

With these data, the form factor is evaluated:
 \beq K_N(t) =
\l\{\begin{array}{lcl}
N&,&t=0\\
|t|/N&,&0<|t|\leq N\\
1&,&N<|t| \end{array}\ri.\qquad (t\in\bZ).
\Leq{fofa}
This calculation is based on the eigenvalues $h_1,\ldots,h_N$ of the unitary
matrix.\footnote{As the $N$ eigenvalues of $U\in U(N)$ have mean distance
$2\pi/N$, note that the natural argument of the form factor would be $t/N$
instead of $t$. However, in order to simplify notation, we use the parameter
$t\in\bZ$.}

In \cite{Be} Berry proposed a semiclassical evaluation of the form
factor for quantum systems, based on the periodic orbits of the
principal symbol (Hamiltonian function) of the Hamiltonian
operator. For the different random matrix ensembles he derived in
the range $0<t\ll N$ the leading order of $K_N(t)$, which is
linear in $t/N$. 

More precisely, semiclassical theory based on the
Gutzwiller trace formula provides a link between spectral
quantities of the quantum Hamiltonian and properties of the
chaotic dynamics of the corresponding classical system. In this
approach the spectral two-point correlation function and its
Fourier transform, namely the form factor, are calculated by
approximating the density of states using the trace formula. 
This formula
expresses them by sums over contributions 
from pairs of classical periodic
trajectories. 

If one includes only pairs of equal or time-reversed
orbits (the so called  ''diagonal
approximation'') then the form factor agrees with random matrix
theory, asymptotically close to the origin (long-range
correlations). 

A more systematic approach will require a complete
control of all the other contributions. A first step towards an
understanding of the ''off-diagonal'' contributions have been
achieved in \cite{BK}. But only recently, beginning with the
article \cite{SR} by Sieber and Richter, contributions involving
pairs of periodic orbits were systematically considered in order
to explain higher order terms in $t/N$.

In particular, for the geodesic flow on constant negative
curvature, a particular family of pairs of periodic orbits have
been presented in \cite{SR} and \cite{Si}, which turned out to be
relevant for the first correction to the diagonal approximation
for the spectral form factor. These orbits pairs are given by
trajectories which exhibit self-intersection with small
intersection angles. This result has been generalized recently to
more general uniformly hyperbolic dynamical systems \cite{Sp}.

The combinatories, however, turned out to be highly nontrivial.
These combinatorial difficulties in handling high order
corrections to the semiclassical expression of the form factor
persist also in the context of quantum graphs, see Kottos and Smilansky
\cite{KS1,KS2},
where these off-diagonal contributions have been explored up to
the third order \cite{Ber1,BSW1,BSW2}.

To our opinion the complex combinatorics should first be studied in the simplest
situation possible, that is, on the group level. Here the unitary ensemble
is the simplest one, since the case of the orthogonal or symplectic ensemble
involves additional elements like the Brauer algebra, see Diaconis and Evans
\cite{DE}.

Now we collect the main points of the article.

We want to compare the form
factor $K_N(t)$ with the diagonal contribution
\beq
\frac{t}{N}\Delta_N^\max(t) \qmbox{with} \Delta_N^\max(t):=\sum_{i_1,\ldots,
i_k=1}^N \LA\prod_{k=1}^t|U_{i_ki_{k+1}}|^2 \RA _N
\Leq{D}
(note that only sum over {\em one} $t$--tuple of indices in $\Delta_N^\max$,
hence the name {\em diagonal contribution}).

The expectation values of products of matrix entries in (\ref{D}) and in
(\ref{KN1}) can be evaluated using the well--known formula (\ref{integration}),
that is, by summing class functions on the symmetric group $S_t$.

So in {\bf Sect.\ \ref{s2}} we introduce some notation concerning the symmetric
group.

In {\bf Sect.\ \ref{s3}} we discuss the relation between the class functions
$V$ and $\cN$ on $\St$ used in (\ref{integration}).
As stated in Prop.\  \ref{prop:Nm1} they are mutual inverses in the group
algebra of $S_t$.

As a simple by--product, this leads to a re--derivation of Eq.\  (\ref{UN})
in the linear regime $|t|\leq N$ (Remarks \ref{rem3.6}).

In {\bf Sect.\ \ref{s4}} we study the relation between a natural metric on
$S_t$ and the joint operation on the associated partition lattice $\cP_t$
(Prop.\  \ref{vee:ineq}).

{\bf Sect.\ \ref{s5}} starts by a (partial) justification of our above
definition (\ref{D}), and an estimate of its contributions in terms of
formula (\ref{integration}).
Here the interplay between the partition lattice and cyclic permutation
becomes essential (Prop.\  \ref{prop:Ct}).
Although Prop.\  \ref{prop:Ct} is a statement about the $N\to\infty$
limit, we present evidence for our conjecture \ref{conj} which is a uniform
in $N\geq t$ version of Prop.\  \ref{prop:Ct}.

In {\bf Sect.\ \ref{s6}} we first prove that only derangements
(that is, fixed point free permutations) are involved in the diagonal
approximation (Prop.\  \ref{prop6.2}).
Then we estimate the number of contributions to $\Delta_N^\max$ with a given
power of $N$ (Prop.~\ref{prop6.3}).

This leads us to our main result in {\bf Sect.\ \ref{s7}}:
Assuming Conjecture \ref{conj}, there exists a subinterval
$I:=[\vep,C-\vep] \subset[0,1]$ such that the diagonal
approximation converges uniformly to the form factor if $t/N\in I$
(Thm.\ \ref{thm}).
\\[2mm]
\noindent
{\bf Acknowledgment:}
This work has been supported by the
European Commission under the Research Training Network
(Mathematical Aspects of Quantum Chaos) no.\ HPRN-CT-2000-00103 of
the IHP Programme.

%
\Section{Generalities on the Symmetric Group}\label{s2}
%
As already mentioned in the Introduction, the symmetric group $\St$
of permutations of the set $[t]:=\{1,\ldots,t\}$ plays an important
r\^{o}le in the analysis of the unitary ensemble.

We begin by introducing some notation, see Sagan \cite{Sag} for
more information. For $\sigma\in \St$ the {\em cycle length} of
$i\in[t]$ is the smallest $n\in\bN$ with $\sigma^n(i)=i$. $i$ is a
{\em fixed point} of $\sigma$ if $n=1$. The {\em cycle} of $i$ is
given by $(i,\sigma(i),\ldots,\sigma^{n-1}(i))$, and can be
interpreted as the group element of $\St$ which permutes the
$\sigma^{k}(i)$ in the prescribed order, leaving the other
elements of $[t]$ fixed.

$e\in\St$ denotes the identity element.

Writing a group element $\sigma\in \St\setminus {\{e\}}$ as a
product of disjoint cycles
$\sigma=\sigma^{(1)}\cdot\ldots\cdot\sigma^{(k)}$, we sometimes
omit the fixed points.

Two lattices are associated with the symmetric group $\St$:
\begin{itemize}
\item
The {\em partition lattice} $\cP_t$ of {\em set partitions} $p=\{a_1,\ldots,
a_k\}$, with {\em atoms} or {\em blocks} $a_l\subseteq[t]$
\big($a_l\cap a_m=\es$
for $l\neq m,\ a_l\neq\es$ and $\bigcup_{l=1}^ka_l=[t]$\big).

$p\in\cP_t$ is called {\em finer} than $q\in\cP_t$ (and $q$ {\em coarser}
than $p$, denoted by $p\preccurlyeq q$) if every block of $p$ is contained
in a block of $q$.

The {\em meet} $p\vee q$ of $p,q\in\cP_t$ is the unique
finest element coarser than $p$ and $q$.

We define the {\em rank} $|p|$ of the partition $p=\{a_1,\ldots,a_k\}\in\cP_t$
by $|p|:=k$ (note that this is called the {\em corank} in \cite{Ai}).
\item
The {\em dominance order} $\Dt$ of {\em number partitions}
$\lambda=(\lambda_1,\ldots,\lambda_k)\in \Dt$ of $t\in\bN$
\big(with $\lambda_l\in\bN,\
\lambda_{l+1}\leq\lambda_l$ and $\sum_{l=1}^k\lambda_l=t$\big). The map
\[\cP_t \ar \cD_t \qmbox{,} \{a_1,\ldots,a_k\}\mapsto(|a_1|,\ldots,|a_k|)\]
induces an order relation and a rank function on $\Dt$.
\end{itemize}
See Aigner \cite{Ai} for more information.

Each permutation $\sigma\in \St$ partitions $[t]$ into atoms belonging to
the same cycle of $\sigma$. Thus we have a map
\[\St\to\cP_t\qmbox{,}\sigma\mapsto
\hat{\sigma}.\]
If the context is clear, we omit the hat. In particular $|\sigma|:=k$
if $\sigma=\sigma^{(1)}\cdot\ldots\cdot\sigma^{(k)}$ is the disjoint
cycle decomposition of $\sigma$ (including fixed points!).
\begin{examples}
\begin{enumerate}
\item
$\sigma=(124)(3)\in S_4$ and $\rho=(142)(3)\in S_4$ have the set partition
$\hat{\sigma}=\hat{\rho}=\{\{1,2,4\},\{3\}\}\in\cP_4$ and number partition
$[\sigma]=[\rho]=(3,1)\in\cD_4$ (Here $[\sigma]:=\{\alpha^{-1}\sigma\alpha
\mid\alpha\in S_t\}$ denotes the conjugacy class of $\sigma\in S_t$).
\item
$\sigma=(12)(34)\in S_4$ and $\rho=(13)(24)\in S_4$ have rank $|\sigma|=|\rho|
=2$, whereas $|\sigma\vee\rho|=|\{\{1,2,3,4\}\}|=1$.
\end{enumerate}
\end{examples}
The importance of the dominance order $\cD_t$ for the symmetric group is
obvious, as the elements of $\cD_t$ naturally enumerate the conjugacy
classes of $\St$. Thus they also enumerate the irreducible representations
and their characters
\[\chi_\lambda:\St\to\bR \qquad (\lambda\in\cD_t).\]
In the present context the importance of the partition lattice
$\cP_t$ comes from the following identity:
\begin{lemma}\label{lem:sum}
For all $t,N,k\in\bN$ and $\pi_1,\ldots,\pi_k\in \St$
\[\sum_{(i_1,\ldots,i_t)\in[N]^t}\prod_{l=1}^k\l(\prod_{j=1}^t\delta_{i_j}^
{i_{\pi_l(j)}}\ri)=N^{|\pi_1\vee\ldots\vee\pi_k|}.\]
\end{lemma}
 From now on our standing assumption relating the groups $\St$ and $U(N)$ is
$t\leq N$.
Then the following important formula can be found in Samuel \cite{Sam},
see also Brouwer and Beenakker \cite{BB}:
\beq
\av{U_{a_1b_1}\ldots U_{a_sb_s}\ov{U}_{\alpha_1\beta_1}\ldots
\ov{U}_{\alpha_t\beta_t}}_N =
\delta_t^s\,\,\sum_{\sigma,\pi\in\St}\V_N(\sigma^{-1}\pi)\,\,\prod_{k=1}^{t}
\delta_{a_k}^{\alpha_{\sigma(k)}}\,\delta_{b_k}^{\beta_{\pi(k)}},
\Leq{integration}
where for $N\geq t$ the class
function $\V\equiv \V_N:\St\ar\bR$
is given by
\beq
\V_N\,:=\,\sum_{\lambda\in\Dt}
\frac{\chi_\lambda(e)}{t! f_\lambda(N)}\ \chi_\lambda.
\Leq{V2}
$f_\lambda$ is a polynomial in $N$ of order $t$ vanishing at certain
integers:
\beq
f_{\lambda}(N):=
\sum_{\sigma\in\St}\frac{\chi_\lambda(\sigma)
N^{|\sigma|}}{\chi_\lambda(e)}
=\prod_{i=1}^k\frac{(N+\lambda_i-i)!}{(N-i)!}\qquad (\lambda\in\Dt)
\Leq{fr}
(see Appendix A of \cite{Sam}).\\[2mm]
Recalling the correspondence between irreducible representations
of $\St$
and conjugacy classes of $\St$, i.e.\ ordered number-partitions
$\lambda=(\lambda_1,\ldots,\lambda_k)\in\Dt$,
$\lambda_1 \geq \ldots \geq \lambda_k$, of $t$, by evaluating Frobenius' Formula
the dimension $\chi_\lambda(e)$ of the representation appearing in (\ref{V2})
equals
$$
\chi_{\lambda}(e)= t!\,
\frac{\prod_{i<j}(\lambda_i-\lambda_j+j-i)}{\prod_i
(\lambda_i+k-i)!}\qquad (\lambda\in\Dt),
$$
see \cite{FH}, Eq. (4.11).

%
%
\Section{The Linear Regime of the Form Factor}\label{s3}
%
Next we decompose the form factor $K_N(t)$ into a sum of products
of the class functions $V_N$ and $\cN:\sigma\mapsto N^{|\sigma|}$ on $\St$.
This will allow us to compare it with the
diagonal contribution to be introduced in Section \ref{diagcon}.

As a side effect, we will re--derive its concrete form (\ref{fofa}) for $|t|\leq N$.
Since $K_N(0)=N^{-1}(\tr(\idty_N))^2=N$ and $K_N(-t)=K_N(t)$, we
effectively only need to consider the regime $0<t\leq N$ where $K_N$ is linear.

Evaluating $\tr(U^t)$ as $\sum_{i\in[N]^t}\prod_{k=1}^tU_{i_ki_{k+1}}$ in
(\ref{KN1}), we get a cyclic ordering of the sub--indices, given by the circular
permutation
\[\tau:=(1,2,\ldots,t)\in \St.\]
Conjugation of $\sigma\in \St$ by $\tau$ will be denoted by $\sigma_+:=
\tau^{-1}\sigma\tau$.

Given $t\in\bN$ and the permutation group $\St$, we denote by
\[\Ct:=\{\sigma\in \St\mid |\sigma|=1\}\]
the set of {\em circular permutations}. This subset is of cardinality
\[|\Ct| =(t-1)!,\]
and every $\sigma\in\Ct$ can be written in the form  $\sigma=\pi^{-1}\tau\pi$
for a unique $\pi\in\St$.
\begin{lemma}\label{lemmino}
The sets
$$
M(\phi,\phi')\,:=\,
\l\{(\pi,\sigma)\in \St\times \St\, \mid\,
\phi=\pi^{-1}\sigma_+
,\,
\phi' = \pi^{-1}\sigma\,\ri\}
\qquad(\phi,\phi'\in\St)
$$
are of size
$$
\left\vert M(\phi,\phi')\right\vert=\left\{\begin{array}{cl} t&
,\ \phi'\tau\phi^{-1}\in\Ct\\0&,\mbox{ otherwise}\end{array}\right.
$$
and form a partition of $\St\times \St$.
\end{lemma}
{\bf Proof:}
\begin{enumerate}
\item
By definition of $M(\phi,\phi')$
any given pair $(\pi,\sigma)\in \St\times \St$ lies in exactly one subset
$M(\phi,\phi')\subseteq\St\times \St$, namely
in
$M(\pi^{-1}\sigma_+,\pi^{-1}\sigma)$.
\item
If $(\pi,\sigma)\in M(\phi,\phi')$ then
$\phi'\tau\phi^{-1}=\pi^{-1}\tau\pi\in \Ct$. Then the $t$ different
\[(\tau^l\pi,\tau^l\sigma)\in\St\times \St\qquad(l=0,\ldots,t-1)\]
are in $M(\phi,\phi')$, too. As thus there are exactly
$\Ct\times \St= (t-1)!\times  t!$
pairs $(\phi,\phi')\in \St\times \St$ with cardinality of the
corresponding atoms
$\left\vert M(\phi,\phi')\right\vert \geq t$, but
$\St\times \St = t!\times t!$, their cardinality must be exactly $t$.
\hfill$\Box$
\end{enumerate}
\begin{proposition}\label{traceprop}
For all $t\leq N\in\bN$, the form factor (\ref{KN1}) equals
\beqn
K_N(t)&=& \frac{t}{N}\cdot
\sum_{\stackrel{\phi,\phi'\in\St}{\phi'\tau\phi^{-1}\in\Ct}}
\V_N(\phi')N^{|\phi|}
\label{eq:K:t}
\\
 &=&  \frac{t}{N}\cdot
\sum_{\gamma\in\Ct}\,\sum_{\phi\in\St}
\V_N(\gamma\phi\tau^{-1}) N^{|\phi|}\,
.
\eeqn
\end{proposition}
{\bf Proof:}
Using sub-indices (mod $t$),
\beqn
\av{\left\vert\tr(U^t)\right\vert^2}_N
&=&\sum_{\stackrel{i_1,\ldots,i_{t}}{j_1,\ldots,j_{t}}}
\av{U_{i_1i_2}\cdots U_{i_{t}i_1}\ov{U}_{j_1j_2}\cdots
\ov{U}_{j_{t}j_1}}_N\nonumber\\
&=&
\sum_{\pi,\sigma\in\St}\V_N(\pi^{-1}\sigma)
\cdot\sum_{\stackrel{i_1,\ldots,i_{t}}{j_1,\ldots,j_{t}}}\prod_{k=1}^{t}
\delta_{i_k}^{j_{\pi(k)}}\cdot\delta_{i_{k+1}}^{j_{\sigma(k+1)}}\nonumber\\
&=&
\sum_{\pi,\sigma\in\St}\V_N(\pi^{-1}\sigma)\cdot
N^{|\pi^{-1}\sigma_+|}\,,
\label{last:step}
\eeqn
since
\beqno
\hspace*{-5mm}
\sum_{\stackrel{i_1,\ldots,i_{t}}{j_1,\ldots,j_{t}}}\prod_{k=1}^{t}
\delta_{i_k}^{j_{\pi(k)}}\cdot\delta_{i_{k+1}}^{j_{\sigma(k+1)}}
&=&\sum_{\stackrel{i_1,\ldots,i_{t}}{j_1,\ldots,j_{t}}}\prod_{k=1}^{t}
\delta_{i_k}^{j_{\pi(k)}}\cdot\delta_{i_{\tau(k)}}^{j_{\sigma\tau(k)}}
=\sum_{\stackrel{i_1,\ldots,i_{t}}{j_1,\ldots,j_{t}}}\prod_{k=1}^{t}
\delta_{i_k}^{j_{\pi(k)}}\cdot\delta_{i_{k}}^{j_{\sigma_+(k)}}\\
&=&
\sum_{\stackrel{i_1,\ldots,i_{t}}{j_1,\ldots,j_{t}}}\prod_{k=1}^{t}
\delta_{i_k}^{j_{\pi(k)}}\cdot\delta_{j_{k}}^{j_{\pi^{-1}\sigma_+(k)}}
=
\sum_{{j_1,\ldots,j_{t}}}\prod_{k=1}^{t}
\delta_{j_{k}}^{j_{\pi^{-1}\sigma_+(k)}}.
\eeqno
In the last step of (\ref{last:step}) we used Lemma
\ref{lem:sum}. Eq.\ (\ref{eq:K:t}) now follows from Lemma \ref{lemmino}.
In (\ref{eq:K:t}) we can write $\phi'=\phi\tau^{-1}\gamma$ for a unique
$\gamma\in \Ct$, This implies the second equation.
\hfill$\Box$\\[2mm]
To further evaluate these expressions for the form factor, we remind the reader
of some general group theoretical notions.

Let $G$ be a finite group with normalized counting measure, that is, the inner
product
\beq
\LA f_1,f_2\RA := \frac{1}{|G|}\sum_{g\in G}\overline{f_1(g)}f_2(g)
\qquad (f_1,f_2\in L^2(G)).
\Leq{inner1}
The characters of the irreducible representations are orthonormal w.r.t.
this inner product and form a basis of the subspace of class functions.

On $L^2(G)$ we have the unitary operators of left and right translations,
given by
\[R_hf(g) := f(gh) \qmbox{,} L_hf(g) := f(hg)\qquad (g,h\in G).\]
For irreducible characters $\chi_\mu,\chi_\lambda:G\to\bC$ one has
\beq
\LA\chi_\lambda,L_g\chi_\mu\RA=\LA\chi_\lambda,R_g\chi_\mu\RA
= \delta_{\lambda\mu}\frac{\chi_\lambda(g)}{\chi_\lambda(e)} \qquad
(g\in G),
\Leq{l1}
see Curtis and Reiner \cite{CR},
Eq.\ (31.16).

We now consider the group algebra $K[\St]$ of the symmetric group,
$K$ denoting a field, i.e.\ the $K$--vectorspace $\{f:\St\ar K\}$,
with multiplication of $f,g\in K[\St]$ given by
\[f*g(\alpha):=\sum_{\sigma\in \St} f(\sigma)g(\sigma^{-1}\alpha)=
\sum_{\sigma\in \St} f(\alpha\sigma^{-1})g(\sigma)
\qquad(\alpha\in \St)\]
and neutral element $\idty_e\in K[\St]$.

More specifically we use the field $K:=\bC(N)$ of rational functions
and denote by ${\cal N}\in K[\St]$ the monomial-valued function
\[{\cal N}(\alpha):=N^{|\alpha|}\qquad(\alpha\in \St)\]
(which, like $\idty_e$, is a class function).

\begin{proposition}\label{prop:Nm1}
$V={\cal N}^{-1}$.
\end{proposition}
{\bf Proof:}
We have, using Eq. (\ref{fr})
\[ V(\phi) =
\sum_{\lambda\in\Dt} \frac{\chi_\lambda(\phi)
\chi_\lambda(e)}{t!f_\lambda(N)}
=
\frac{1}{t!}\sum_{\lambda\in\Dt}
\frac{\chi_\lambda(\phi)(\chi_\lambda(e))^2}
{\sum_{\sigma\in \St}\chi_\lambda(\sigma)N^{|\sigma|}}
\]
Thus (as $|\sigma^{-1}|=|\sigma|$) we must prove that \beq
\sum_{\lambda\in\Dt} \frac{\sum_{\omega\in
\St}\chi_\lambda(\alpha\omega)N^{|\omega|}} {\sum_{\sigma\in
\St}\chi_\lambda(\sigma)N^{|\sigma|}} (\chi_\lambda(e))^2 =
t!\idty_e(\alpha). \Leq{CC} In order to show that the l.h.s.\ is
in fact independent of $N$ (if $N\geq t$ so that the
denominator does not vanish!), for $\sigma\in \St$ we sum over the
conjugacy class $[\sigma]\subseteq \St$ of $\sigma$, using that
$|\rho\sigma\rho^{-1}| = |\sigma|$.

More specifically we claim the existence of a constant $C_\lambda(\alpha)$
such that for all $\sigma\in \St$
\beq
\sum_{\tilde{\sigma} \in [\sigma]}
\chi_\lambda(\alpha\tilde{\sigma})
= C_\lambda(\alpha)\sum_{\tilde{\sigma}\in[\sigma]}\chi_\lambda(\tilde{\sigma}).
\Leq{BB}
Equivalently we show that
\[{\LL}_\lambda(\sigma)\equiv{\LL}_\lambda(\alpha,\sigma) :=
\sum_{\rho\in \St}\chi_\lambda
(\alpha\rho\sigma\rho^{-1})\]
equals $C_\lambda(\alpha)\cdot r_\lambda(\sigma)$ with
$r_\lambda(\sigma) := \sum_{\rho\in \St}\chi_\lambda(\rho\sigma
\rho^{-1}) = t!\chi_\lambda(\sigma)$.

Now
for $\alpha\in \St$
\beqno
{\LL}_\lambda(\sigma) &=& \sum_{\rho\in \St}\chi_\lambda
((\rho^{-1}\alpha\rho)\sigma)\\
&=&\frac{|\St|}{|[\alpha]|}
\sum_{\pi\in[\alpha]}\chi_\lambda(\pi\sigma)
=\frac{t!}{|[\alpha]|}
\sum_{\pi\in[\alpha]}L_{\pi}\chi_\lambda(\sigma).
\eeqno
${\LL}_\lambda$ being a class function, we write it in the form
\beq
{\LL}_\lambda = \sum_{\mu\in\Dt}d_\mu\chi_\mu
\Leq{AA}
and determine the coefficients $d_\mu$ using the orthonormality
relation $\LA\chi_\lambda,\chi_\mu\RA = \delta_{\lambda\mu}$.
By Eq.\ (\ref{l1})
\[\LA L_{\pi}\chi_\lambda,\chi_\mu\RA = \delta_{\lambda\mu}
\frac{\chi_\lambda(\pi)}{\chi_\lambda(e)}\]
which leads to
\[d_\mu = \delta_{\lambda\mu}\frac{t!}{|[\alpha]|\chi_\lambda(e)}
\sum_{\pi\in[\alpha]}
\chi_\lambda(\pi).\]
Inserting this into (\ref{AA}) we see that $C_\lambda(\alpha)$ in (\ref{BB})
equals
\[C_\lambda(\alpha) = \frac{1}{|[\alpha]|\chi_\lambda(e)}
\sum_{\pi\in[\alpha]}
\chi_\lambda(\pi).\]
Using:
$$
\sum_{\sigma\in\St}\chi_{\lambda}(\alpha \sigma)N^{|\sigma|}\,=
\,C_{\lambda}(\alpha)\,\,\sum_{\sigma\in\St}
\chi_{\lambda}(\sigma)N^{|\sigma|}\, ,
$$
the l.h.s.\ of (\ref{CC}) equals
\beqn
\lefteqn{\frac{1}{|[\alpha]|}\sum_{\lambda\in\Dt}\sum_{\pi\in[\alpha]}
\chi_\lambda(\pi)\chi_\lambda(e)}\nonumber\\
 &=&\frac{1}{|[\alpha]|}\sum_{\pi\in[\alpha]}\sum_{\lambda\in\Dt}
\chi_\lambda(\pi)\chi_\lambda(e)=
\frac{1}{|[\alpha]|}\sum_{\pi\in[\alpha]}
 \idty_e(\pi)\sum_{\lambda\in\Dt}
(\chi_\lambda(e))^2\nonumber\\
&=& \frac{1}{|[\alpha]|}\idty_e(\alpha)
t! = \idty_e(\alpha)t!,
\label{chi:square}
\eeqn
using the identity
$$
\sum_{\lambda\in\Dt}\chi_{\lambda}(e)^2\,=\,t !
$$
in (\ref{chi:square}), see  Chapter 5.2 in \cite{Sag} and Rains \cite{Ra}.
This  proves (\ref{CC}).
\hfill $\Box$\\[2mm]
We redefine the inner product (\ref{inner1})
on $\bC[\St]$ omitting the factor $1/|\St|=1/t!$ :
\beq
\LA f_1,f_2\RA := \sum_{\sigma\in \St}\overline{f_1(\sigma)}f_2(\sigma)
\qquad (f_1,f_2:S_t\ar\bC).
\Leq{inner2}
So the irreducible characters are now of norm $t!$. Anyhow we are
now more interested in the following sets of functions:

Instead of considering the field $\bC(N)$ of rational functions
in the variable $N$ we will now
specialize the value $N\in\bN$, $N\geq t$.

Define for $\sigma\in \St$ the translates of $\cN$:
\[\hq_\sigma\in \bC[S_t]\qmbox{,}\hq_\alpha(\sigma) := N^{|\sigma^{-1}\alpha|}
\qquad (\alpha\in S_t).\]

Similarly we define the translates
\[V_\sigma\in\bC[S_t]\qmbox{,} V_\alpha(\sigma) := V_N(\sigma^{-1}\alpha)
\qquad (\alpha\in S_t)\]
of $V_N$.

\begin{lemma}
\label{lem:basis}
For $N\geq t$ the $\hq_\sigma$, $\sigma\in S_t$ form a basis of the vector space
$\bC[S_t]$, with dual basis $V_\sigma$, $\sigma\in S_t$.
\end{lemma}
{\bf Proof:}
Considered as rational functions, for $\alpha,\beta\in\St$
the inner product equals
\beqno
\LA V_\alpha, \hq_\beta\RA &=&
\sum_{\sigma\in \St}V_N(\sigma^{-1}\alpha)N^{|\sigma^{-1}\beta|}
=\sum_{\sigma\in \St}V_N(\alpha^{-1}\sigma)N^{|\sigma^{-1}\beta|}\\
 &=&
\sum_{\rho\in \St}V_N(\rho)N^{|\rho^{-1}(\alpha^{-1}\beta)|}=
V*\cN(\alpha^{-1}\beta)=\delta_\alpha^\beta.
\eeqno
Specializing the value of $N$, this duality relation is true as long as
the rational functions are defined.
By inspection of the definition (\ref{V2}) of $V_N$ (in particular of the
$f_\lambda$ defined in (\ref{fr})) this is the case as long as $N\geq t$.
As the number of the  $V_\alpha$ and of the $\hq_\beta$ both
equals $\dim(\bC[S_t])=t!$, these are indeed bases.
\hfill$\Box$
\begin{corollary} \label{prop:CC}
For $t\leq N$
$$
\sum_{\sigma\in\St} V_N(\alpha\sigma^{-1})N^{|\sigma|} =
\idty_e(\alpha) \qquad(\alpha\in\St).
$$
\end{corollary}
\begin{remarks}\label{rem3.6}
\begin{enumerate}
\item
Corollary \ref{prop:CC} allows us to regain formula (\ref{fofa}), i.e.
\[K_N(t) = \frac{t}{N} \qmbox{for} 0<t\leq N.\]
Using Prop.\  \ref{traceprop} we have
\beqno
K_N(t) & = & \frac{t}{N}\sum_{\gamma\in C_t}\sum_{\phi\in \St}V_N(\gamma\phi\tau^{-1})N^{|\phi|}\\
&=& \frac{t}{N}\sum_{\gamma\in C_t}\sum_{\sigma\in \St}
V_N(\sigma)N^{|\gamma^{-1}\sigma\tau|} =
\frac{t}{N}\sum_{\gamma\in C_t}\idty_\gamma(\tau) = \frac{t}{N}.
\eeqno
\item
As $V_N:S_t\to\bR$ is a class function, we can also use the notation
\[V_N:\cD_t\to\bR\qmbox{,} V_N([\sigma]):= V_N(\sigma).\]
Then we can calculate $V_N$ using Prop.\  \ref{prop:Nm1}. Some examples:
\begin{itemize}
\item
for $t=1$ we have $V_N(1)=\frac{1}{N}$;
\item
for $t=2$ and denominator $D_2 := N(N^2-1)$ we have
\[V_N(1,1)=\frac{N}{D_2}\qmbox{,} V_N(2)=-\frac
{1}{D_2};\]
\item
for $t=3$ and $D_3 := N^3(N^2-1)(N^2-4)$ we have
\[V_N(1,1,1)=\frac{N^4-2N^2}{D_3}\qmbox{,} V_N(2,1)=-\frac{N^3}{D_3}\qmbox{and}
V_N(3)=\frac{2N^2}{D_3}.\]
\end{itemize}
\item
The large $N$ asymptotics of $V_N:\cD_t\to\bR$ for $\lambda=(\lambda_1,\ldots,
\lambda_k)$ is given by
\beq
V_N(\lambda)\sim(-1)^{t-k}N^{k-2t}\prod_{l=1}^kC_{\lambda_l} \qquad (N\to\infty)
\Leq{Bee}
with the Catalan number $C_l:={2l-2\choose l-1}/l$, see \cite{Sam}.
\end{enumerate}
\end{remarks}
%

\Section{The Rank Function and the Join of Partitions}\label{s4}
%

It is useful to give a geometric meaning to our estimates.
For this purpose we equip the symmetric group $\St$ with
the metric
\[d:\St\times \St\ar\{0,1,\ldots,t\}\qmbox{,}
d(\sigma,\gamma) = t-|\sigma\gamma^{-1}|.\]
The easiest way to visualize this metric is to consider the ${t\choose2}$--regular
{\em Cayley graph}
$(\St,E_t)$ having the symmetric group as its vertex set, and edge set
\[E_t :=  \l\{(\rho,\rho')\in \St\times \St\mid
\rho^{-1}\rho'\mbox{ is a transposition }\ri\}.\]
\begin{proposition} \label{vee:ineq}
\begin{enumerate}
\item
$d(\sigma,\gamma)$ is the distance between the vertices $\sigma$ and
$\gamma$ on the Cayley graph $(\St,E_t)$.
So in particular the metric $d$ is invariant under the left and right
self-actions of $\St$.
\item
$\qquad\qquad|\pi\vee\sigma|=\min\,\big\{|\pi\mu^{-1}| \mid \mu\preccurlyeq\sigma\big\}
\qquad(\pi,\sigma\in \St)$.\\
So in particular
\[d(\pi,\sigma)\leq t-|\pi\vee\sigma|.\]
\item
$\qquad\qquad d(\rho,\rho')\geq
\l\vert\,\, |\rho\vee\sigma|-|\rho'\vee\sigma |\,\,\ri\vert
\qquad(\rho,\rho',\sigma\in \St)$.
\end{enumerate}
\end{proposition}
{\bf Proof:}
\begin{enumerate}
\item
For $\rho := \sigma\gamma^{-1}$ with disjoint cycle decomposition
$\rho=\rho_1\cdot\ldots\cdot\rho_k$ we have
$d(\sigma,\gamma)=d(\rho,e)=t-k=\sum_{i=1}^k (l_i-1)$, $l_i$ being the length
of $\rho_i$.
Exactly $l-1$ transpositions are needed to form a cycle of length $l$.
\item
Let $(c_1,\ldots, c_m)$, $c_k\subseteq\{1,\ldots,t\}$ be the partition
corresponding to the cycles of $\pi$.
We consider the graph $(V,E)$ with vertex set $V := \{c_1,\ldots, c_m\}$
and edges $\{c_i,c_j\}\in E$ for which there are elements
$e_i\in c_i$, $e_j\in c_j$ which belong to the same cycle of $\sigma$.

Choose  for each connected component of $(V,E)$ a spanning tree and
representatives $\{e_i,e_j\}$ of its edges. Then by construction
the product $\mu_0$
of the transpositions $(e_i,e_j)$ meets $\mu_0\preccurlyeq\sigma$,
and $|\pi\mu_0^{-1}|=|\pi\vee\sigma|$. Any $\mu\preccurlyeq\sigma$
can be written in the form $\mu=\rho\mu_0$ with $\rho\preccurlyeq\sigma$.
As no cycles of $\pi\mu_0^{-1}$ can be joined by right multiplication with
$\rho^{-1}\preccurlyeq\sigma$, the statement follows.
\item
By symmetry of the metric $d$ we assume
$|\rho\vee\sigma|\geq|\rho'\vee\sigma |$ and choose $\mu_0\preccurlyeq\sigma$
so that $|\rho'\vee\sigma|=|\rho'\mu_0^{-1}|$. Then, again by Part 2 of the proposition
\[0\leq|\rho\vee\sigma|-|\rho'\vee\sigma |\leq |\rho\mu_0^{-1}|- |\rho'\mu_0^{-1}|.\]
By Part 1 of the proposition
\[ |\rho\mu_0^{-1}|- |\rho'\mu_0^{-1}|\leq
d(\rho\mu_0^{-1},\rho'\mu_0^{-1})=d(\rho,\rho')\]
since multiplication by a transposition changes the number of cycles by one,
and since $d$ is invariant under right multiplication.\hfill $\Box$
\end{enumerate}
\begin{remark}
As the elements $\rho=\sigma=(12)(34)$, $\rho'=(13)(24)$ of $S_4$ show,
in general the inequality
$|\,|\rho\vee\sigma|-|\rho'\vee\sigma |\,|\leq|\,|\rho|-|\rho'|\,|$ does
{\bf not} hold.
The reverse inequality is wrong, too in general.
\end{remark}
%
\Section{The Diagonal Contribution}\label{diagcon}\label{s5}
%
We now define and study the {\it diagonal approximation} for the
unitary ensemble.

Setting $[N]  := \{1,\ldots,N\}$, the diagonal contribution is defined
by:
\beq
\Delta_N(t):= \sum_{i\in [N]^t} {\rm per}(i)
\LA \prod_{k=0}^{t-1} \left\vert
U_{i_ki_{k+1}}\right\vert^2\RA,
\Leq{dia}
where ${\rm per}(i)$ denotes
the period of $i$.
In fact (see (\ref{integration})), only those terms of the form
factor
\[K_N(t)\,=
\sum_{i,j\in [N]^t}\LA\prod_{k=0}^{t-1}
U_{i_ki_{k+1}}\ov{U}_{j_kj_{k+1}}\RA\]
can be non-zero for which the
sets
\[m_i(r):=\{k\in[t]\mid i_k=r\}\]
have equal multiplicity ($|m_i(r)|=m_i(r)$ for all $r\in[N]$).

In this case, if only multiplicities $|m_i(r)|\leq1$ occur
for $i$, there is a unique permutation $\sigma$ with
$j_{\sigma(k)}=i_k$ (and $\pi := \tau\sigma\tau^{-1}$ with
$j_{\pi(k)+1}=i_{k+1}$), but in general we have
 \[K_N(t)\,=
\sum_{i,j\in [N]^t} \sum_{\kappa^{(1)},\kappa^{(2)}\in
S(m;(1))\times\ldots\times S(m;(N))}
V\l((\sigma\kappa^{(1)})^{-1}\pi\kappa^{(2)}\ri).\]
As the dominant (in
$N$) contributions are the ones with $V(e)$, i.e.\ $\sigma=\tau^l$
for some $l$, we call the sum

\begin{eqnarray}
\Delta_N(t) &=& \sum_{i,j\in [N]^t\exists l \tiny{\mbox{ with }}
j_{k+l}=i_k } \LA \prod_{k=0}^{t-1}
U_{i_ki_{k+1}}\ov{U}_{j_kj_{k+1}}\RA\\
&&\nonumber\\
&&\nonumber\\
&=&  \sum_{i\in [N]^t} {\rm per}(i) \LA \prod_{k=0}^{t-1}
\left\vert U_{i_ki_{k+1}}\right\vert^2\RA. \end{eqnarray}
the
{\em diagonal contribution}.

If $|m_i(r)|\leq 1$, then only terms $V(e)$ occur in $\LA
\prod_{k=0}^{t-1} \left\vert U_{i_ki_{k+1}}\right\vert^2\RA$.

The number of all terms in $\sum_{i_1,\ldots,i_t}$ being $N^t$,
for $k|t$
\[I_k := \{(i_1,\ldots,i_t)\mid i_{l+k}=i_l\}\]
is the set of terms with ${\rm per}(i_1,\ldots,i_t)|k$.
So the number of terms with\newline
${\rm per}(i)<t$ equals
\[ \sum_{r>1,r|t} |I_{t/r}| \mu(r),\]
with the M\"obius $\mu$ function. As $|I_k|=N^k$, this is only of
order $N^{t/2}\log(t)$ and thus negligible compared to $|I_t|=N^t$
as $N\ar\infty$.

For these reasons, we just define and study a function similar to
(\ref{dia}) but replacing the period by its maximal value $t$.
In fact for simplicity of notation we use the constant one instead:
$$
\Delta^{\rm max}_N(t)\, := \,\sum_{i_1,\ldots,i_t}\brac{\prod_{k=0}^{t-1}
\left\vert U_{i_ki_{k+1}}\right\vert^2}
$$
A basic manipulation yields:
\begin{proposition}
\beq
\Delta^{\rm max}_N(t)\,=\,\sum_{\pi,\sigma\in\St} \V_N(\pi^{-1}\sigma)
N^{\left\vert\pi\vee\sigma_+\right\vert},
\Leq{dia:max}
\end{proposition}
{\bf Proof:}
Using (\ref{integration}),
\begin{eqnarray}
\Delta^{\rm max}_N(t)&=&\sum_{i\in[N]^t}\sum_{\pi,\sigma\in\St}
\V_N(\pi^{-1}\sigma)\cdot\prod_{k=0}^{t-1}
\delta_{i_k}^{i_{\pi(k)}}\cdot\delta_{i_{k+1}}^{i_{\sigma(k+1)}}\nonumber\\
&=&\sum_{i\in[N]^t}\sum_{\pi,\sigma\in\St}
\V_N(\pi^{-1}\sigma)\cdot\prod_{k=0}^{t-1}
\delta_{i_k}^{i_{\pi(k)}}\cdot\delta_{i_{k}}^{i_{\tau^{-1}\sigma\tau(k)}}.
\end{eqnarray}
Lemma  (\ref{lem:sum}) now gives the result.
\hfill$\Box$\\[2mm]
A first easy observation is that for {\em bounded} $t$

$$\label{easyone}
\Delta^{\rm max}_N(t)\,=\, 1+\cO(1/N^2).
$$
This follows by inserting (\ref{Bee}) into (\ref{dia:max}).
\begin{remark}
In general for $0<t\leq N$ {\bf neither} $K_N(t)=\frac{1}{N}\Delta_N(t)$
{\bf nor} $K_N(t)=\frac{t}{N}\Delta_N^\max(t)$, although both equations hold
for $N=1$ and $N=2$.
Already for $t=3\leq N$
\[\Delta_N(3)=\frac{3N^3}{D_3}(N^4-7N^2+4N+2)\neq3=NK_N(3)\]
and
\[\Delta_N^\max(3)=\frac{N^3}{D_3}(N^4-3N^2-6N+8)\neq1=N\frac{K_N(3)}{3},\]
with denominator $D_3=N^3(N^2-1)(N^2-4)$.

So the diagonal approximation is not exact.
\end{remark}
According to (\ref{Bee}) the terms in the sum (\ref{dia:max}) have fluctuating
sign:
\[\sign\l(\V_N(\pi^{-1}\sigma)\ri)=(-1)^{d(\pi,\sigma)}.\]
This makes it advisable to perform a partial summation before estimating
terms in absolute value.
We thus rewrite the sum over $\pi$ in (\ref{dia:max}) in the form of an inner
product:
\beq
\Delta^{\rm max}_N(t)\,=\,\sum_{\sigma\in\St} \LA\V_N,\hp_\sigma\RA\qmbox{with}
\hp_\sigma(\alpha) :=
N^{|\sigma\alpha^{-1}\vee\sigma_+|}.\Leq{one:sum:only}

\begin{proposition}\label{prop:Ct}
There exists a function $C_t:\St\to\{0,1\}$ such that
\[\LA V_N,\hp_\sigma\RA_N
= N^{|\sigma\vee\sigma_+|-t}(C_t(\sigma)+\cO(1/N))\qquad (\sigma\in \St).\]
\end{proposition}
{\bf Proof:}
For $\sigma\in \St$ the symmetric group is partitioned into the sets
\[B_n := \{\alpha\in \St\mid|\alpha\vee\sigma_+| = |\sigma_+|-n\} \qquad
(n=0,\ldots,|\sigma|-1).\]
The metric $d$ on $\St$ is then used to introduce for $\gamma\in B_n$
\[B(\gamma) := \l\{\alpha\in\bigcup_{k=0}^nB_k\mid|\alpha\vee\sigma_+| -
|\gamma\vee\sigma_+|=d(\alpha,\gamma)\ri\}.\] Observe that by Part
3 of Lemma \ref{vee:ineq} we always have \beq
0\leq|\alpha\vee\sigma_+|-|\gamma\vee\sigma_+|\leq
d(\alpha,\gamma). \Leq{Y} In particular $\gamma$ is the only
element in $B(\gamma)\cap B(n)$. This enables us to define for
$n=0,\ldots,|\sigma|-1$ \beq C_t(\gamma) := 1-\sum_{\alpha\in
B(\gamma)\backslash\{\gamma\}}C_t(\alpha) \qquad (\gamma\in B_n),
\Leq{X} and the approximants
\[\tilde{p}_\sigma:\St\to\bR \qmbox{,} \tilde{p}_\sigma := \sum_{\gamma\in \St}
C_t(\gamma)\, N^{|\gamma\vee\sigma_+|-t}\widehat{q}_{\gamma^{-1}\sigma}
\qquad (\sigma\in \St)\] of the functions $\hp_\sigma$.
\begin{itemize}
\item
Next we prove that $C_t$ only takes the values $0$ and $1$. This follows
from the definition (\ref{X}), if we can show that each $\gamma$ has exactly
one {\em predecessor} in
\[P := \{\alpha\in \St\mid C_t(\alpha)=1\},\]
that is, $|B(\gamma)\cap P|=1$. This is done by induction in $n$,
with
\[\gamma\in P\cap B_n \qquad (n=0,\ldots,|\sigma|-1)\]
and noting that $P\cap B_0=B_0$ (the $\gamma\in B_0$ are their own predecessors
so that $C_t(\gamma)=1$).
\item
For the induction step we use the directed graph $(\St,E)$ with vertex set
$\St$ and edges
\[(\alpha,\beta)\in E\Leftrightarrow d(\alpha,\beta)=1\ \mbox{and}\ \alpha
\in B_n,\beta\in B_{n-1}\ \mbox{for\ some}\
n\in\{1,\ldots,|\sigma|-1\}.\] By the triangle inequality for
$\gamma\in B(n)$ the set $B(\gamma)$ contains all $\alpha\in B_k,\
0\leq k\leq n$ for which there exists a directed chain
\[\gamma=c_n,c_{n-1},\ldots,c_k=\alpha\ \mbox{from}\ \gamma\ \mbox{to}\ \alpha\
\mbox{with}\ c_l\in B_l\ \mbox{and}\ (c_l,c_{l-1})\in E\]
$(l=k+1,\ldots,n)$.

Conversely all $\alpha\in B(\gamma)$ are of that form. Namely for $\alpha\in
B(\gamma)\cap B_k$ we know that $d(\alpha,\gamma) = n-k$ so that there exist
$c_n,\ldots,c_k\in \St$ with $c_k=\gamma,\ c_k=\alpha$ and $d(c_{l-1},c_l)=1$.
As $||c_{l-1}\vee\sigma_+|-|c_l\vee\sigma_+||\leq d(c_{l-1},c_l)=1$ and
$|c_n\vee\sigma_+|=n,\ |c_k\vee\sigma_+|=k$, we conclude $|c_l\vee\sigma_+|
=l$ so that $(c_l,c_{l-1})\in E$.
\item
This shows that $\alpha\in P$ if there does not exist an edge $(\alpha,\beta)
\in E$, and thus $|B(\gamma)\cap P|\geq1$ (as every directed chain starting
at $\gamma$ ends somewhere).

To prove that $|B(\gamma)\cap P|=1$, we need a more precise
characterization of the predecessors $\alpha\in P$. As there does
not exist an edge of the form $(\alpha,\beta)$ in $E$, for all
neighbors $\beta\in S_+$ of $\alpha$ (i.e. $d(\alpha,\beta)=1$) we
have $|\beta\vee\sigma_+|\leq|\alpha\vee\sigma_+|$. In other words
if $\beta$ differs from $\alpha$ by a transposition, and if two
blocks of $\hat{\sigma}_+\in P_t$ belong to the same block of
$\alpha\vee \sigma_+$, then they belong to the same block of
$\beta\vee\sigma_+$. \item We model this by considering for given
$\sigma\in \St$ the directed multigraph
\[G_\alpha = (V_\alpha,\cE_\alpha)\]
associated to $\alpha\in \St$. The vertex set of $G(\alpha)$ equals
$V_\alpha := \{\hat{\sigma}_{+,1},\ldots,\hat{\sigma}_{+,m}\}$, with
$\hat{\sigma}_{+,1},\ldots,\hat{\sigma}_{+,m}\subseteq[N]$ the blocks of the
set partition $\hat{\sigma}_+\in P_t$.

The multiplicity of the directed edge $(\hsi_{+,i},\hsi_{+,j})\in V_\alpha
\times V_\alpha$ is given by
\[\cE_\alpha:V_\alpha\times V_\alpha\setminus \Delta\to\bN_0\ ,\
\cE_\alpha(\hsi_{+,i}, \hsi_{+,j}) :=
\big|\{(u,v)\in\hsi_{+,i}\times\hsi_{+,j}\mid\alpha(u)=v\}\big|.\]
The in- and outdegrees of the blocks $\hsi_{+,i}$ coincide,
that is
$\cE_\alpha^+(\hsi_{+,i})=\cE_\alpha^-(\hsi_{+,i})$ for
\[\cE_\alpha^+(\hsi_{+,i}) := \sum_{\hsi_{+,j}}\cE_\alpha(\hsi_{+,i},\hsi_{+,j})
\qmbox{,} \cE_\alpha^-(\hsi_{+,i}) :=
\sum_{\hsi_{+,j}}\cE_\alpha(\hsi_{+,i},\hsi_{+,j}).\] Henceforth
we omit the superscripts $\pm$ and simply refer to the {\em
degree} $\cE_\alpha(\hsi_{+,i})=\cE_\alpha^\pm(\hsi_{+,i})$ of the
block. \item All ancestors $\alpha\in P$ have multigraphs
$G_\alpha$ which have two--connected components, that is, the
number of connected components cannot be in\-creased by reducing a
single degree $\cE(\hsi_{+,i})$ by one. This can be seen by
noticing that for every $\alpha\in \St$ the number of connected
components of $G_\alpha$ equals $|\alpha\vee\sigma_+|$, and using
that the $\alpha\in P$ don't have neighbors $\beta$ with
$|\beta\vee\sigma_+| =|\alpha\vee\sigma_+|+1$.
\item
To prove $|B(\gamma)\cap P|=1$, we assume
that $\alpha^{(1)},\alpha^{(2)}\in B(\gamma)\cap P$. So there
exist directed chains $\gamma=c_n^{(i)},c_{n-1}^{(i)},\ldots,
c_{k^{(i)}}^{(i)}=\alpha^{(i)}$ (with
$\l(c_l^{(i)},c_{l-1}^{(i)}\ri)\in E$) from $\gamma$ to
$\alpha^{(i)},\ i=1,2$, and we are to show that $\alpha^{(1)}=
\alpha^{(2)}$. In each step the number
$|c_l^{(1)}\vee\sigma_+|=|c_l^{(2)}\vee \sigma_+|=l$ of connected
components of the multigraphs $G_{c_l^{(i)}}$ is reduced by one.
That is, all connected components of the multigraph $G_\gamma$ are
broken into their two--connected subcomponents:
\[\cE_{\alpha^{(i)}}\leq\cE_\gamma \qmbox{and} \cE_{\alpha^{(i)}}
(\hsi_{+,j})\neq1 \qquad (i=1,2,\ j=1,\ldots,m).\] In fact this
shows that $\cE_{\alpha^{(1)}}=\cE_{\alpha^{(2)}}$ so that the
multigraphs of $\alpha^{(1)}$ and $\alpha^{(2)}$ coincide.

The multiplicity $\cE_\gamma(\hsi_{+,i}\hsi_{+,j})$ of a directed
edge of $G_\alpha$ is reduced only if
$\cE_\gamma(\hsi_{+,i},\hsi_{+,j})=1$. So not only
$\cE_{\alpha^{(1)}}(\hsi_{+,i},\hsi_{+,j})=\cE_{\alpha^{(2)}}
(\hsi_{+,i},\hsi_{+,j})$ but the chains connecting $\gamma$ with
$\alpha^{(1)} =\alpha^{(2)}$. \item We now know that $C_t(\alpha)$
only takes the values $0$ and $1$, and that
$\tilde{p}_\sigma=\sum_{\gamma\in P}\,N^{|\gamma\vee\sigma_+|-t}
\hq_{\gamma^{-1}\sigma}$. This implies
\[\LA V_N,\tilde{p}_\sigma\RA_N = \sum_
{\gamma\in P}\,N^{|\gamma\vee\sigma_+|-t}\LA
V_N,\hq_{\gamma^{-1}\sigma}\RA_N =
C_t(\sigma)N^{|\sigma\vee\sigma_+|-t}.\]

It remains to show that
\[\LA V_N,\hp_\sigma\RA_N = \LA V_N,\tilde{p}_\sigma\RA_N+\cO(N^{|\sigma\vee
\sigma_+|-t-1}).\]
But, denoting the unique predecessor of $\beta\in \St$ by $P(\beta)$
(that is $\{P(\beta)\} = B(\beta)\cap P$), we have
\beqno
\tilde{p}_\sigma(\beta^{-1}\sigma) &=&
 \sum_{\gamma\in P}N^{|\gamma\vee\sigma_+| -d(\gamma,\beta)}\\
&=& N^{|P(\beta)\vee\sigma_+|-d(P(\beta),\beta)}+
\sum_{\stackrel{\gamma\in P}{\gamma\neq P(\beta)}}
N^{|\gamma\vee\sigma_+|-d(\gamma,\beta)}. \eeqno By definition of
$B(\gamma)$ the exponent of the first term equals
\[\big|P(\beta)\vee\sigma_+\big|-
d\big(P(\beta),\beta\big)=|\beta\vee\sigma_+|,\]
whereas the exponents of the second term are smaller:
\[|\gamma\vee\sigma_+|-d(\gamma,\beta)<|\beta\vee\sigma_+|\]
on the other hand
\[\hp_\sigma(\beta^{-1}\sigma)=N^{|\beta\vee\sigma_+|},\]
proving the claim.
\hfill $\Box$
\end{itemize}
If $\sigma\in\St$ consists of a single nontrivial cycle,
the estimate of Prop.\  \ref{prop:Ct} can be
replaced by an identity (Prop.\  \ref{propo:cycle} below).

We prepare this by a sum rule for the class function $\cN$:

\begin{lemma}
\label{lem:sum:SkN}
For all $k\in\bN$
\beq
\sum_{\sigma\in S_k}N^{|\sigma|} = \prod_{l=0}^{k-1}(N+l).
\Leq{C}
\end{lemma}
{\bf Proof:}
For $k=1$ both sides equal $N$. So assume the formula to
hold for $k-1$, so that \beq \sum_{\tilde{\sigma}\in
S_k,\tilde{\sigma}(k)=k}N^{|\tilde{\sigma}|} =
N\prod_{l=0}^{k-2}(N+l). \Leq{A} The group elements $\sigma\in
S_k$ either have $k$ as a fixed point $s$ or can uniquely be
written in the form
\[\sigma = (l,k)\tilde{\sigma}\]
with $l\in\{1,\ldots,k-1\}$ and $\tilde{\sigma}(k)=k$. As in the
second case $|\sigma|=|\tilde{\sigma}|-1$, \beq
\sum_{l=1}^{k-1}\sum_{\tilde{\sigma}\in S_k,\tilde{\sigma}(k)=k}
N^{|\tilde{\sigma}|} = (k-1)\prod_{l=0}^{k-2}(N+l). \Leq{B} Adding
the contributions (\ref{A}) and (\ref{B}) yields (\ref{C}).
\hfill $\Box$\\[2mm]
We now decompose $\hp_\sigma$ in the form
\[\hp_\sigma = \sum_{\gamma\in \St}c_\gamma\hq_\gamma \qmbox{with}
c_\gamma := \LA V_\gamma,\hp_\sigma\RA.\]
\begin{proposition} \label{propo:cycle}
For a cycle $\sigma = (i_1+1,\ldots,i_k+1)\in \St$ 
(and $\sigma_+  = (i_1,\ldots,i_k)$)
\beq
\hp_\sigma =
\l(\prod_{l=1}^{k-1}(N+l)\ri)^{-1}\sum_{\gamma'\preccurlyeq\sigma_+}
\hq_{\gamma'\sigma}.
\Leq{exact}
\end{proposition}
{\bf Proof:}
$\bullet$
We evaluate both sides on $\tilde{\alpha}\in \St$ and write
$\tilde{\alpha}$ as $\tilde{\alpha} = \alpha\sigma$ to simplify
expressions. Then
\beq
\hp_\sigma(\alpha\sigma) = N^{|\alpha^{-1}\vee\sigma_+|} \qmbox{and}
\hq_{\gamma'\sigma}(\alpha\sigma) = N^{|\alpha^{-1}\gamma'|}.
\Leq{pq}
$\bullet$
Next we write $\alpha$ as a product of disjoint cycles $z_j$ and
note that
\[|z_j\alpha^{-1}\vee\sigma_+|-|z_j\alpha^{-1}\gamma'| = |\alpha^{-1}
\vee\sigma_+|-|\alpha^{-1}\gamma'|\]
if $z_j$ and $\sigma_+$ are
disjoint. We thus can reduce 
$\alpha$ to a product of cycles intersecting $\sigma_+$.\\
$\bullet$
So we assume w.l.o.g.\ that all cycles $z_j$ of $\alpha$
intersect $\sigma_+= (i_1,\ldots,i_k)$:
\[z_j = \l(i_{\pi(1)},\tilde{z}_1,\ldots,\tilde{z}_2,i_{\pi(2)},\tilde{z}_3,
\ldots,\tilde{z}_{2s-2},i_{\pi(s)},\tilde{z}_{2s-1},\ldots,\tilde{z}_{2s}\ri)\]
with $\tilde{z}_n\in\{1,\ldots t\}\backslash\{i_1,\ldots,i_k\}$.
Then \beqno \l(i_{\pi(s)},i_{\pi(s-1)},\ldots,i_{\pi(1)}\ri)z_j
&=& \l(i_{\pi(1)},\tilde{z}_1,\ldots,\tilde{z}_2\ri)
\l(i_{\pi(2)},\tilde{z}_3,\ldots,\tilde{z}_4\ri)\ldots\\
&&\l(i_{\pi(s)},\tilde{z}_{2s-1},\ldots,\tilde{z}_{2s}\ri) 
\eeqno
is a product of disjoint cycles intersecting the cycle $\sigma_+$
only at $i_{\pi(n)}$. Furthermore
\[\l(i_{\pi(s)},i_{\pi(s-1)},\ldots,i_{\pi(1)}\ri)\preccurlyeq\sigma_+\]
so that the map
\[\gamma'\mapsto\l(i_{\pi(s)},i_{\pi(s-1)},\ldots,i_{\pi(1)}\ri)\gamma'\]
simply permutes the $\gamma'$ in
$\sum_{\gamma'\preccurlyeq\sigma_+}\hq_{\gamma' \sigma}$.
This allows to reduce to the case of simple intersections.\\
$\bullet$
We thus assume w.l.o.g.\ that the cycles $z_j$ in the
decomposition of $\alpha$ intersect $\sigma_+$ exactly in one
point, say $i_j$. Under this assumption, by Lemma
\ref{lem:sum:SkN} and (\ref{pq})
\beqno \lefteqn{\l(\prod_{l=1}^{k-1}(N+l)\ri)^{-1}
\sum_{\gamma'\preccurlyeq\sigma_+}\hq_{\gamma'\sigma}(\alpha\sigma)}\\
&=& \l(\prod_{l=1}^{k-1}(N+l)\ri)^{-1}N^{|\alpha^{-1}|-k}
\sum_{\gamma\in S_k}N^{|\gamma|}=
N^{|\alpha^{-1}|-k+1}\\
&=& N^{|\alpha^{-1}\vee \sigma_+|}=\hp_\sigma(\alpha\sigma),
\eeqno proving the assertion.
\hfill$\Box$
\begin{corollary}\label{corr}
For a cycle $\sigma = (i_1,\ldots,i_k)\in \St$ of length
$k$
\beq
\LA V_N,\hp_\sigma\RA=\l\{\begin{array}{cl}1&,\, k=1 \qmbox{that is} \sigma=e\\
0&,\, 1<k<t\\
\l(\prod_{l=1}^{t-1}(N+l)\ri)^{-1}&,\, k=t\end{array}\ri.
\ .
\Leq{cycle:zero}
\end{corollary}
{\bf Proof:}
This follows from Prop.\  \ref{propo:cycle} with
$\LA V_N,\hq_\sigma\RA=\delta_{e,\sigma}$ (Lemma \ref{lem:basis}), 
remarking that only for
$k=1$ or $k=t$ there is a $\gamma'\preccurlyeq\sigma_+$ with
$\gamma'\sigma= e$.
\hfill$\Box$\\[2mm]
This result and numerical experiments support the following
conjecture (compare with Prop.\  \ref{prop:Ct}):
\begin{conjecture}\label{conj}
There exists a constant $C_1\geq 1$ such that for all $t\leq N\in\bN$
\[\l|\LA V_N,\hp_\sigma\RA_N\ri| \leq
C_1 N^{|\sigma\vee\sigma_+|-t}\qquad (\sigma\in \St).\]
\end{conjecture}

%

%
\Section{Derangements and Circular Order}\label{s6}
%
We now show that, apart from the identity, only the {\em derangements},
that is the fixed-point free permutations
\[D_t := \{\sigma\in S_t\mid\sigma(k)\neq k
\qmbox{for all} k\in\{1,\ldots,t\}\}\]
contribute in the sum (\ref{one:sum:only}).

This will follow from a statement of independent interest:
\begin{proposition}\label{prop6.1}
For $k=1,\ldots,t+1$ denote by  $\St^{(k)}$ the subgroup
\[\St^{(k)}:=\{\sigma\in S_{t+1}\mid \sigma(k)=k\},\]
and by $I_k:\St\ar \St^{(k)}$ the isomorphism induced by the injection
\[\tilde{I}_k: \{1,\ldots,t\}\hookrightarrow\{1,\ldots,t+1\}\qmbox{,}
\tilde{I}_k(i)=\l\{\begin{array}{cl}
i&,\, 1\leq i<k\\ i+1&,\, i\geq k.\end{array}\ri. .\]
Then for $\sigma=I_{k+1}(\tilde{\sigma})$ and
\[\hp_{\tilde{\sigma}}=
\sum_{\tilde{\gamma}\in\St} c_{\tilde{\gamma}}\,  \hq_{\tilde{\gamma}\tilde{\sigma}}\]
we have
\[\hp_\sigma=
\sum_{\tilde{\gamma}\in\St} c_{\tilde{\gamma}}\, \hq_{I_k(\tilde{\gamma})\sigma}.\]
\end{proposition}
{\bf Proof:}
$\bullet$
For $\beta\in \St^{(k)}\subset S_{t+1}$, that is $\beta=I_k(\tilde{\beta})$
with $\tilde{\beta}\in\St$
\beqno
\hp_\sigma \l( \beta \sigma \ri) &=&
N^{\l| I_k(\tilde{\beta}) \vee \sigma_+ \ri|}
= N^{\l| I_k(\tilde{\beta}) \vee (I_{k+1}(\tilde{\sigma}))_+ \ri|}
=   N^{\l| I_k(\tilde{\beta}) \vee I_{k}(\tilde{\sigma}_+) \ri|}\\
&=& 
N^{\l| \tilde{\beta} \vee \tilde{\sigma}_+ \ri|+1}=
N \hp_{\tilde{\sigma}}(\tilde{\beta}\tilde{\sigma})
\eeqno
and similarly
\beqno
\hq_{I_k(\tilde{\gamma})\sigma}\l(\beta\sigma \ri) &=&
N^{|I_k(\tilde{\gamma})(I_k(\tilde{\beta}))^{-1}|}
= N^{|I_k(\tilde{\gamma}\tilde{\beta}^{-1})|} =
N^{|\tilde{\gamma}\tilde{\beta}^{-1}|+1}\\
&=& N \hq_{\tilde{\gamma}\tilde{\sigma}}(\tilde{\beta}\tilde{\sigma}).
\eeqno
$\bullet$
The other elements of $S_{t+1}$ can be uniquely written as
a product of a  transposition $(l,k)\in S_{t+1}$  and
$\beta=I_k(\tilde{\beta})\in \St^{(k)}$.
In that case a similar argument leads to
\[\hp_\sigma ( \beta \sigma ) =
\hp_{\tilde{\sigma}}(\tilde{\beta}\tilde{\sigma})\qmbox{and}
\hq_{I_k(\tilde{\gamma})\sigma}\l(\beta\sigma \ri) =
\hq_{\tilde{\gamma}\tilde{\sigma}}(\tilde{\beta}\tilde{\sigma}).\]
$\bullet$
So in any case the proportionality factor does not depend on $\tilde{\gamma}$.
\hfill$\Box$
\begin{proposition}\label{prop6.2}
For all $t\leq N\in\bN$
\beq
\LA\V_N,\hp_\sigma\RA = 0
\qmbox{for} \sigma\in S_t\setminus D_t \, ,\,\sigma\neq e.
\Leq{condition}
\end{proposition}
{\bf Proof:}
Lemma \ref{lem:basis} implies the formula
\[\LA\V_N,\hq_\sigma\RA =\delta_e^\sigma.\]
So (\ref{condition}) is equivalent to show that for these $\sigma$ in the
base decomposition
\[\hp_\sigma=\sum_{\gamma\in\St} c_{\gamma} \hq_{\gamma\sigma}\]
of $\hp_\sigma$ the coefficient $c_{\sigma^{-1}}$ equals zero.
These $\sigma$ have a fixed point $k+1$ (mod $t$) which has the additional
property that $k$ (mod $t$) is {\em not} a fixed point.
So $\sigma=I_{k+1}(\tilde{\sigma})$ with $\tilde{\sigma}\in S_{t-1},\ \tilde
{\sigma}(k)\neq k$.
The base decomposition $\hp_{\tilde{\sigma}}=\sum_{\tilde{\gamma}\in S_{t-1}}
c_{\tilde{\gamma}}\hq_{\tilde{\gamma}\tilde{\sigma}}$ leads to $\hp_\sigma=
\sum_{\tilde{\gamma}\in S_{t-1}}c_{\tilde{\gamma}}\hq_{{I_k}(\tilde{\gamma})
\sigma}$, see Prop.\  \ref{prop6.1}.

Thus if the $\tilde{\gamma}\in S_{t-1}$
term in (\ref{condition}) would be non--zero, it would be of the form
$\LA V_N,\hp_\sigma\RA = c_{\tilde{\gamma}}$ for $\tilde{\gamma}\in S_{t-1}$ with
$I_k(\tilde{\gamma})=\sigma^{-1}=I_{k+1}(\tilde{\sigma}^{-1})$ or
$I_{k+1}(\tilde{\sigma})=I_k(\tilde{\gamma}^{-1})$. But this would imply
$\tilde{\sigma}(k)=k$, contradicting the assumption.
\hfill$\Box$\\[2mm]
It is known that
\[|D_t|\sim\frac{|\St|}{e}\qmbox{as} t\to\infty.\]
So it could seem that we would only gain an unimportant factor $1/e$
by restricting the summation in (\ref{one:sum:only}) to the derangements
(and the identity).

This is not so, since we can use the structure
of the derangements under the $\tau$ action in our estimation.

For that purpose we now partition the derangements $D_t$ by setting
\[D_t(k) := \{\sigma\in D_t\mid|\sigma \vee\sigma_+| = k\} \qquad
(k = 1,\ldots,t).\]
So $D_t(k) = \es$ for $k>t/2$, and we estimate the
cardinalities of these sets.
\begin{proposition} \label{prop6.3}
There exists a $C_2\geq 1$ such that for all
$t\in\bN$
\[|D_t(k)|\leq k C_2^k (t-k+1)!\qquad(k = 1,\ldots,\lfloor t/2\rfloor).\]
\end{proposition}
{\bf Proof:}
Remark that the statement becomes trivial for $k = 1$
so that in the proof we assume $k\geq2$.

\begin{itemize}
\item
Each $\sigma\in D_t(k)$ induces a set partition
\[B(\sigma) \equiv B = (B_1,\ldots,B_k)\]
of $\{1,\ldots,t\}$ into the blocks of $\sigma\vee\sigma_+$ which
is unique if you assume $|B_{l+1}| \geq |B_l|$ and
$\min(B_l)\leq\min(B_{l+1})$ if $|B_{l+1}| = |B_l|$. As each $B_l$
contains at least one cycle of $\sigma$ (or rather the block
corresponding to the cycle in the partition of $\sigma$), we have
$|B_l| \geq 2$.
\item
Next we consider the intersections
\[C_{l,m} := B_l\cap B_m^+ \qquad (l,m\in\{1,\ldots,k\})\]
with the atoms $B_m^+ := \tau(B_m) = \{j+1\mid j\in B_m\}$ of the
shifted set partition $B^+$. We thus get a set partition
\[C(\sigma) \equiv C = (C_{1,1},\ldots,C_{k,k})\]
of $\{1,\ldots,t\}$ which is finer than $B$ and $B^+$ but may
contain empty atoms $C_{l,m}$. However, as $\sigma$ is a
derangement, we know that if $C_{l,m}(\sigma)$ is nonempty, it is
a union of cycles of $\sigma$ so that in any case
$|C_{l,m}(\sigma)|\neq1$.
\item
We now estimate $|D_t(k)|$ by
\[|D_t(k)|\leq\sum_{b = (b_1,\ldots,b_k)}Y(b)\]
where $2\leq b_1\leq\ldots\leq b_k,\ \sum_{l=1}^kb_l = t$ and
\[Y(b) := |\{\sigma\in D_t\mid|B_l(\sigma)| = b_l,\ l = 1,\ldots,k\}|.\]
\item
This quantity, in turn is estimated by \beq
Y(b)\leq\sum_{c = (c_{1,1},\ldots,c_{k,k})}X(c)\prod_{l,m=1}^kc_{l,m}!
\Leq{1} where now $c_{l,m}\in\{0,\ldots,b_l\}\backslash\{1\}$ with
$\sum_{m=1}^k c_{l,m} = b_l$ and
\[X(c) = \big|\big\{B = (B_1,\ldots,B_k)\mid|C_{l,m}| =
c_{l,m}\big\}\big|.\]

Here $\{B_1,\ldots,B_k\}$ is an arbitrary set partition of
$\{1,\ldots,t\}$ with enumeration fixed by demanding
$2\leq|B_1|\leq\ldots\leq|B_k|$ and, again,
$\min(B_l)\leq\min(B_{l+1})$ if $|B_l| = |B_{l+1}|$. Denoting as
before by $C_{l,m}$ the intersection $B_l\cap B_m^+$ formula
(\ref{1}) follows by our above remark that all $\sigma\in D_t$
with $C_{l,m}(\sigma) = C_{l,m}$ have a cycle partition finer than
$C = (C_{1,1},\ldots,C_{k,k})$ and there are $c_{l,m}!$ ways to
permute the set $C_{l,m}$.

We bound $X(c)$ by considering the directed multigraph $G = G(c)$
with vertex set $V := \{1,\ldots,k\}$ and $c_{l,m}$ unlabeled
directed edges from vertex $l$ to $m$. Then
\beq
X(c)\leq X_G(c),
\Leq{2}
where $X_G(c)$ is the number of closed Euler trails on
$G$. This can be seen as follows:
\begin{enumerate}
\item
The length of any closed Euler trail equals $\sum_{l,m=1}^k
c_{l,m} = t$.
\item
Any closed directed Euler trail on $G$
(shortly called {\em trail} from now on) is uniquely characterized
by the sequence $(v_1,\ldots,v_t)$ of vertices $v_j\in V$ it
visits. This is due to our assumptions that the edges from $l$ to
$m$ are unlabeled, and that the beginning of the closed trail is
marked.
\item
A set partition $B = (B_1,\ldots,B_k)$ of
$\{1,\ldots,t\}$ gives rise to a sequence $(v_1,\ldots,v_t)$ of
vertices $v_i\in V$, where $v_i := j$ if $i\in B_j$. Using a
$t$--periodic notation with $v_{t+1} = v_1$, we have
\[|\{i\in\{1,\ldots,t\}\mid(v_i,v_{i+1}) = (l,m)\}| = c_{l,m} \qquad
(l,m\in\{1,\ldots,k\}).\]
Thus $B$ gives rise to a trail in $G(c)$
rooted at $v_1\in V$.
\end{enumerate}
It may be remarked that we have equality in (\ref{2}) if the
vertices of the directed multigraph $G(c)$ can be discerned by
their outdegree, that is $b_1<\ldots<b_k$. Then, given an Euler
trail with sequence $(v_1,\ldots,v_t)$, we define the partition
$(B_1,\ldots,B_k)$ by setting $B_j := \{i\in\{1, \ldots,t\}\mid
v_i = j\}$.
\item
To get an upper bound on $X_G(c)$ we select a root
vertex $j\in V$ and consider the Euler trails in $G(c)$ beginning
at $j$. By the BEST formula their number equals \beq
b_j^RT_j(c)\cdot\frac{\prod_{l=1}^k(b_l-1)!}{\prod_{l,m=1}^kc_{l,m}!}
\Leq{3} where $T_j(c)$ is the number of directed spanning trees
rooted at $j$. (\ref{3}) is derived from Thm. 13 of Chapter I of
\cite{Bo} by noting that, unlike here, Bollobas considers directed
multigraphs with labeled edges. Here the {\em reduced outdegree}
\[b_l^R := \sum_{m=1}^kc_{l,m}^R \qmbox{with} c_{l,m}^R := 1 \mbox{ if }
c_{l,m}>0 \mbox{ and } c_{l,m}^R := 0 \mbox{ otherwise.}\]
\item
The number of directed spanning trees rooted at $j$ equals the
$(k-1)\times(k-1)$--minor of the $k\times k$ degree matrix
\[{\rm diag}(b_1^R,\ldots,b_k^R) - (c_{l,m}^R)_{l,m=1}^k\]
obtained by deleting the $j$--th row and the $j$--th column. This
number is known to be independent of $j$, and we call it
$\Delta(c^R)$.

By this remark and (\ref{3}) \beq X_G(c)\leq
t\Delta(c^R)\frac{\prod_{l=1}^k (b_l-1)!}{\prod_{l,m=1}^k
c_{l,m}!}, \Leq{4} since $\sum_{j=1}^k b_j^R \leq
\sum_{j=1}^kb_j=t$.
\item
 From (\ref{1}) and (\ref{4}) we obtain the estimate
\beq Y(b)\leq
t\prod_{l=1}^k(b_l-1)!\cdot\sum_{c=(c_{1,1},\ldots,c_{k,k})}
\Delta(c^R). \Leq{5} A bound on $\Delta(c^R)$ only depending on
the reduced outdegrees $b_1^R,\ldots,b_k^R$ can be found in
\cite{GM}. We use it in the slightly weakened version
\[\Delta(c^R)\leq\eh\prod_{l=1}^k b_l^R\]
and thus get from
(\ref{5}) \beq
Y(b)\leq\textstyle{\frac{t}{2}}\prod_{l=1}^k\l[b_l^R(b_l-1)!\ri]\cdot\sum_c1.
\Leq{6}
\item
The cardinality $\sum_c1$ of number partitions
$c = (c_{1,1},\ldots,c_{k,k})$ compatible with the number partition
$(b_1,\ldots,b_k)$ of $t$ is calculated as follows:
\[\sum_c1 = \prod_{l=1}^k\l|\l\{(c_{l,1},\ldots,c_{l,k})\mid c_{l,m}\neq1\
\mbox{and}\ \sum_{m=1}^kc_{l,m} = b_l\ri\}\ri|.\]
But
\beqno
\lefteqn{\l|\l\{(c_{l,1},\ldots,c_{l,k})\mid c_{l,m}\neq1\
\mbox{and}\
\sum_{m=1}^kc_{l,m} = b_l\ri\}\ri|}\\
&=&
\sum_{U\subseteq\{1,\ldots,k\}}\Big|\Big\{(c_{l,1},\ldots,c_{l,k})\mid
c_{l,m} \geq 2\ \mbox{if}\ m\in U\ \mbox{and}\ c_{l,m} = 0 \\
&& \hspace{58mm}\mbox{otherwise,}\ \sum_{m\in U}c_{l,m} = b_l\Big\}\Big|\\
&=& \sum_{r=1}^{\min(k,\lfloor b_l/2\rfloor)}\sum_{|U|=r}\bsm
b_l-r-1\\ r-1 \esm\\
&=& \sum_{r=1}^{\min(k,\lfloor b_l/2\rfloor)}\bsm b_l-r-1\\ r-1
\esm \bsm k\\ r\esm, \eeqno so that (\ref{6}) reduces to \beq
Y(b)\leq\frac{t}{2}\prod_{l=1}^k\l[b_l^R(b_l-1)!\sum_{r=1}^{\min(k,\lfloor
b_l/2\rfloor)} \bsm b_l-r-1\\ r-1 \esm \bsm k\\ r\esm\ri]. \Leq{7}
\item
We bound the sums appearing in (\ref{7}), depending on the
relative size of $k$ and $b_l$. Remember our assumption $k \geq 2$.

We set $\hat{b}: = \lfloor b/2\rfloor$.
\begin{enumerate}
\item
For all $k,b \geq 2$ we have the estimate \beqn\label{8}
\lefteqn{\sum_{r=1}^{\min(k,\hat{b})} \bsm b-r-1\\ r-1 \esm \bsm
k\\ r \esm}\nonumber\\
&\leq& \sum_{r=1}^{\min(k,b-1)} \bsm b-r-1\\ r-1 \esm \bsm k\\ r
\esm \leq \sum_{r=1}^{\min(k,b-1)} \bsm b-2\\ r-1 \esm \bsm
k\\ k-r \esm \nonumber\\
&=& \bsm k+b-2\\ k-1 \esm \eeqn
\item
For all $k\geq b \geq 2$ and
$r\leq\hat{b}$ we use the inequality $\bsm k\\ r \esm\leq \bsm k\\
\hat{b} \esm \leq \bsm k\\ \lfloor k/2\rfloor \esm$ to show
\beqn\label{9} \sum_{r=1}^{\min(k,\hat{b})} \bsm b-r-1\\ r-1 \esm
\bsm k\\ r \esm &\leq& \l(\sum_{r=1}^{\hat{b}} \bsm b-r-1\\ r-1
\esm\ri)\bsm
k\\ \hat{b} \esm\nonumber\\
&=& \frac{g^{b-1}-\l(\frac{-1}{g}\ri)^{b-1}}{\sqrt{5}} \bsm k\\
\hat{b} \esm \eeqn
with the golden mean $g := \frac{1+\sqrt{5}}{2}$, since the sum of
the binomials equals the Fibonacci numbers.
\end{enumerate}
\item
The reduced outdegree $b_l^R$ is bounded by \beq b_l^R =
\sum_{m=1}^k c_{l,m}^R\leq\min(k,\hat{b}_l)\leq\frac{2k\hat{b}_l}
{k+\hat{b}_l}. \Leq{10} Instead of summing (\ref{7}) over the
ensemble of $b = (b_1,\ldots,b_k)$ with $2\leq b_1\leq\ldots\leq
b_k$ and $\sum_{l=1}^kb_l = t$, we shift the $b_l$ by $2$ and set
for $\tilde{k}\leq k$
\[Y_k(b_1,\ldots,b_{\tilde{k}};t) := \frac{t}{2}k^{k-\tilde{k}}\prod_{l=1}
^{\tilde{k}}b_l^R(b_l+1)!\sum_{r=1}^{\min(k,\hat{b}_l)} \bsm
b_l-r+1\\ r-1 \esm \bsm k\\ r \esm,\]
with $\hat{b}_l :=
\l\lfloor\frac{b_l}{2}\ri\rfloor+1$ and $b_l^R$ is redefined as
$\min(k,\hat{b}_l)$.

Then for $b_l \geq 0$ \beq Y_G(b_1+2,\ldots,b_k+2)\leq
Y_k(b_1,\ldots,b_k;t) \Leq{12} and our aim is to find a $C \geq 1$,
independent of $k$ and $t$, such that the recursion \beq
\sum_{\stackrel{0\leq b_1\leq\ldots\leq
b_{\tilde{k}+1}}{\sum_{l=1}^{\tilde{k}+1}b_l=t-2k}}
Y_k(b_1,\ldots,b_{\tilde{k}+1};t)\leq C\cdot \sum_{\stackrel{0\leq
b_1\leq\ldots\leq b_{\tilde{k}}}{\sum_{l=1}^{\tilde{k}} b_l=t-2k}
} Y_k(b_1,\ldots,b_{\tilde{k}};t) \Leq{11} in $\tilde{k}$ holds
true. Assuming (\ref{11}), we obtain from (\ref{12})
\[\sum_{\stackrel{2\leq b_1\leq\ldots\leq b_k}{\sum_{l=1}^kb_l=t}}
Y_G(b_1,\ldots,b_k)\leq k e(c e)^{k-1}(t-k+1)!,\]
since \beqno
\lefteqn{Y_k(t-2k;t)\leq\frac{t}{2}k^{k-1}\min\l(k,\l\lfloor\frac{t}{2}-k\ri
\rfloor+1\ri)(t-2k+1)!\bsm k+t-2k\\ k-1 \esm}\\
&=&
\frac{t}{2}\frac{k^k}{k!}\min\l(k\l\lfloor\frac{t}{2}-k\ri\rfloor+1\ri)
(t-k)!\\
&\leq& ke^k(t-k+1)!. 
\eeqno 
Using (\ref{8}), (\ref{11}) follows
from the recursion 
\beqno
\lefteqn{\sum_{l=0}^{\hat{b}-1}\hat{l}(l+1)!\l(\sum_{r=1}^{\min(k,\hat{l})}
\bsm l-r+1\\ r-1 \esm \bsm k\\r \esm\ri) \widehat{b-l}(b-l+1)!
\bsm k+b-l\\ k-1 \esm }\\
&\leq& C\cdot k(\hat{b})(b+1)!\bsm k+b\\ k-1 \esm, \eeqno with
$\hat{l} = \l\lfloor\frac{l}{2}\ri\rfloor+1$. this is equivalent
to the claim \beq
\sum_{l=0}^{\hat{b}-1}\hat{l}(l+1)!\l(\sum_{r=1}^{\min(k,\hat{l})}
\bsm l-r+1\\ r-1 \esm \bsm k\\ r \esm\ri) \widehat{b-l}
\prod_{r=1}^l \frac{1}{k+b-r+1} \leq Ck\hat{b}. \Leq{13} Depending
on the relative size of $k$ and $b$, we estimate the l.h.s.\ of
(\ref{13}) in two ways:
\item
 Using (\ref{8}), we get the uppper
bound for the l.h.s.\ of (\ref{13}) \beqn\label{14}
\lefteqn{\sum_{l=0}^{\hat{b}-1}\hat{l}(l+1)! \bsm k+l\\ k-1 \esm
\widehat{b-l}\prod_{r=1}^l \frac{1}{k+b-r+1}}\nonumber\\
&=& k\sum_{l=0}^{\hat{b}-1}\hat{l}\widehat{b-l}\prod_{r=1}^l\frac
{k+r}{k+b-r+1}. \eeqn We write the product in (\ref{14}) in the
form
\[\prod_{r=1}^l\frac{k+r}{k+b-r+1} = \exp\l(\sum_{r=1}^lg(r)\ri)
\qmbox{with} g(r) := \ln \l(\frac{k+r}{k+b-r+1}\ri).\]
For $r\leq
l\leq\widehat{b-1}$ not only $g(r)\leq0$ and $g(r)\geq g(r-1)$ but
also (for all real such $r$)
\[g''(r) = \frac{1}{(k+b-r+1)^2} - \frac{1}{(k+r)^2}\leq0.\]
So $\sum_{r=1}^lg(r)\leq lg\l(\frac{1+l}{2}\ri)\leq
lg\l(\frac{\hat{b}}{2}\ri)$ or \beq
\prod_{r=1}^l\frac{k+r}{k+b-r+1}\leq\lambda^l \qmbox{with} \lambda
:= \frac{k+\frac{\hat{b}}{2}}{k+b-\frac{\hat{b}}{2}+1}. \Leq{15}
For $k\leq b$ we have the uniform bound $\lambda\leq\frac{3}{4}$.

Inserting (\ref{15}) in (\ref{14}) and noting that
$\widehat{b-l}\leq \hat{b}$ and $\hat{l}\leq\frac{l}{2}+1$, we get
(\ref{13}) with
\[C := \sum_{l=0}^\infty\l(\frac{l}{2}+1\ri)\l(\frac{3}{4}\ri)^l=10.\]
\item
For $k\geq b$ we insert (\ref{9}) in the l.h.s.\ of
(\ref{13}) which is thus bounded by \beqno
\lefteqn{\sum_{l=0}^{\hat{b}-1}\hat{l}(l+1)!g^l\bsm k\\ \hat{l}-1
\esm
\widehat{b-l}\prod_{r=1}^l\frac{1}{k+b-r+1}}\\
&=& \sum_{l=0}^{\hat{b}-1}\hat{l}\widehat{b-l}g^l \bsm k\\
\hat{l}-1 \esm \prod_{r=1}^l\frac{r+1}{k+b-r+1}. \eeqno By an
argument similar to the one leading to (\ref{15}) \beqno
\lefteqn{\sum_{l=0}^{\hat{b}-1}\hat{l}\widehat{b-l}g^l \bsm k\\
\hat{l}-1 \esm
\prod_{r=1}^l\frac{r+1}{k+b-r+1}}\\
&=& \sum_{l=0}^{\hat{b}-1}\hat{l}^2\widehat
{b-l}\prod_{r=1}^{\lfloor
l/2\rfloor}\frac{g^2(r+\hat{l})}{(k+b-r+1)
(k+b-r-\lfloor l/2\rfloor+1)}\\
&\leq&
\sum_{l=0}^{\hat{b}-1}\hat{l}^2\widehat{b-l}\prod_{r=1}^{\lfloor
l/2 \rfloor}\frac{g^2}{k+b-r+1} \leq
\hat{b}\sum_{l=0}^{\hat{b}-1}\hat{l}^2
\l(\frac{g}{2}\ri)^l\\
&\leq& \hat{b}\sum_{l=0}^\infty\l(\frac{l^2}{4}+l+1\ri)
\l(\frac{g}{2}\ri)^l = \frac{1}{4}(161+71\sqrt{5})<80, \eeqno
assuming $k\geq 4$ and treating $k=2$ and $k=3$ separately. \hfill
$\Box$
\end{itemize}
%
\Section{The Asymptotic Estimate}\label{s7}
%
Now we are ready to present our asymptotic result.
\begin{theorem}\label{thm}
Under the assumption of Conjecture \ref{conj} the form factor $K_N$ is
approximated by the diagonal contribution in the following sense:\\
For all $\vep>0$ uniformly in $\frac{t}{N}\in\l[\vep,\frac{e}{C_2}(1-\vep)\ri]$
\[\l|K_N(t)-\frac{t}{N}\Delta_N^\max(t)\ri|\to0 \qquad (N\to\infty).\]
\end{theorem}
{\bf Proof:}
As $K_N(t)=\frac{t}{N}$ for the $t$--values under consideration,
\[\Delta_N^\max(t)=\sum_{\sigma\in S_t}\LA V_N,\hp_\sigma\RA\]
(Eq. (\ref{one:sum:only})), and $\LA V_N,\hp_e\RA=1$ (Cor. \ref{corr}), we
need to show that
\[\sum_{\sigma\in S_t\backslash\{e\}}\LA V_N,\hp_\sigma\RA\to0\qquad (N\to\infty).\]
Using Prop.\  \ref{prop6.2} this amounts to show
\[\sum_{\sigma\in D_t}\LA V_N,\hp_\sigma\RA\to0 \qquad (N\to\infty),\]
which is implied by the asymptotic vanishing of
\[\sum_{k=1}^{\lfloor t/2\rfloor}\sum_{\sigma\in D_t(k)}|\LA V_N,\hp_\sigma\RA|
\leq C_1\sum_{k=1}^{\lfloor t/2\rfloor}kC_2^k(t-k+1)!N^{k-t},\]
using Conjecture \ref{conj} and Prop.\  \ref{prop6.3}. Under our assumptions
for $t/N$
\beqno
\lefteqn{\sum_{k=1}^{\lfloor t/2\rfloor}kC_2^k(t-k+1)!N^{k-t}\leq tN\sum_{k=1}^{\lfloor
t/2\rfloor}C_2^k\l(\frac{t-k+1}{N^e}\ri)^{t-k+1}}\\
&\leq& N^2\sum_{k=1}^{\lfloor t/2\rfloor}C_2^k\l(\frac{1-\vep}{C_2}\ri)^
{t-k+1}\\
&\leq& N^2C_2^{-1}\sum_{k=1}^{\lfloor t/2\rfloor}(1-\vep)^{t-k+1}\\
&\leq& \frac{N^2}{C_2\vep}\cdot(1-\vep)^{\lceil t/2\rceil+1}\leq\frac{N^2}
{C_2\vep}(1-\vep)^{\vep N/2}\to0.
\eeqno
proving the theorem.
\hfill $\Box$
%

\addcontentsline{toc}{section}{References}
\end{document}